\documentclass[prd,superscriptaddress,showpacs,nofootinbib,amsmath,amssymb,aps,10pt]{revtex4}

\usepackage{bm}
\usepackage{amsfonts}
\usepackage{latexsym}
\usepackage[latin1]{inputenc}
\usepackage{graphicx}
\usepackage{amsmath}
\usepackage{palatino}
\usepackage{mathpazo}
\usepackage[outdir=./]{epstopdf}

\usepackage{booktabs}
\usepackage{dcolumn}

\def\jnl@style{\it}
\def\aaref@jnl#1{{\jnl@style#1}}

\def\aaref@jnl#1{{\jnl@style#1}}

\def\aj{\aaref@jnl{AJ}}                   
\def\apj{\aaref@jnl{ApJ}}                 
\def\apjl{\aaref@jnl{ApJ}}                
\def\apjs{\aaref@jnl{ApJS}}               
\def\apss{\aaref@jnl{Ap\&SS}}             
\def\aap{\aaref@jnl{A\&A}}                
\def\aapr{\aaref@jnl{A\&A~Rev.}}          
\def\aaps{\aaref@jnl{A\&AS}}              
\def\mnras{\aaref@jnl{Mon.~Not.~Roy.~Astron.~Soc.}}             
\def\prd{\aaref@jnl{Phys.~Rev.~D}}        
\def\prc{\aaref@jnl{Phys.~Rev.~C}}  
\def\prl{\aaref@jnl{Phys.~Rev.~Lett.}}    
\def\qjras{\aaref@jnl{QJRAS}}             
\def\skytel{\aaref@jnl{S\&T}}             
\def\ssr{\aaref@jnl{Space~Sci.~Rev.}}     
\def\zap{\aaref@jnl{ZAp}}                 
\def\nat{\aaref@jnl{Nature}}              
\def\aplett{\aaref@jnl{Astrophys.~Lett.}} 
\def\apspr{\aaref@jnl{Astrophys.~Space~Phys.~Res.}} 
\def\physrep{\aaref@jnl{Phys.~Rep.}}      
\def\physscr{\aaref@jnl{Phys.~Scr}}       
\def\commat{\aaref@jnl{Comm.~Math.~Phys.}}              
\def\science{\aaref@jnl{Science}}               
\def\cqg{\aaref@jnl{Classical Quant.~Grav.}}            
\def\jpcs{\aaref@jnl{JPCS}}                                     
\def\ijmpd{\aaref@jnl{Int.~J.~Mod.~Phys.~D}}                    
\def\grg{\aaref@jnl{Gen.~Relat.~Gravit.}}               
\def\rpp{\aaref@jnl{Rep.~Prog.~Phys.}}          
\def\npa{\aaref@jnl{Nucl.~Phys.~A}}        
\def\lrr{\aaref@jnl{Living Rev.~Rel.}}                   
\def\jcap{\aaref@jnl{J.~Cosmology Astropart.~Phys.}}    
\def\rmp{\aaref@jnl{Rev.~Mod.~Phys.}}   

\allowdisplaybreaks[1]

\begin{document}

\title{Comment on ``The Mass-Radius relation for Neutron Stars in $f(R)$ gravity'' \\ by S. Capozziello, M. De Laurentis, R. Farinelli and S. Odintsov}
\author{Stoytcho S. Yazadjiev}
\email{yazad@phys.uni-sofia.bg}
\affiliation{Department
	of Theoretical Physics, Faculty of Physics, Sofia University, Sofia
	1164, Bulgaria}
\affiliation{Theoretical Astrophysics, Eberhard-Karls University
	of T\"ubingen, T\"ubingen 72076, Germany}

\author{Daniela D. Doneva}
\email{daniela.doneva@uni-tuebingen.de}
\affiliation{Theoretical Astrophysics, Eberhard-Karls University
	of T\"ubingen, T\"ubingen 72076, Germany}
\affiliation{INRNE - Bulgarian Academy of Sciences, 1784  Sofia, Bulgaria}


\begin{abstract}
In the present comment we discuss the qualitative and the quantitative differences between the neutron stars models in $f(R)$ gravity presented in the recent paper  \cite{Capozziello2015} (Capozziello et al. (2015)) and in our paper \cite{Yazadjiev2014} (Yazadjiev et al. (2014)). The authors of \cite{Capozziello2015} claim that these differences come from the fact that they are using the physical Jordan frame while we are working in the Einstein frame. The results in \cite{Yazadjiev2014} are very well tested by two numerical codes using different formulations of the field equations that give us the confidence that they are correct. Thus we decided to perform an additional test. Namely, we derive the relevant set of field equations in the Jordan frame and solve them numerically in order to disprove the claim in \cite{Capozziello2015} that the difference comes from the use of different frame. The obtained results are in  perfect agreement with our previous work \cite{Yazadjiev2014} and therefore the numerical results in \cite{Capozziello2015} seem to be incorrect. At the end we point out several possible sources of error in \cite{Capozziello2015} that can lead to the observed large deviations.
\end{abstract}

\maketitle
\section{Introduction}

In the paper \cite{Capozziello2015} the authors study equilibrium models of static neutron stars in $f(R)$ gravity and in particular in the $R$-squared gravity. The approach is non-perturbative and they adopt the Jordan frame. Their results are qualitatively and quantitatively different from the previous studies of neutron stars in  $R$-squared gravity reported in our paper \cite{Yazadjiev2014} where the mathematical equivalence between $f(R)$ theories and scalar-tensor theories is used. The authors of \cite{Capozziello2015} suggest   that this is due to the fact that they use the physical Jordan frame, while the studies in \cite{Yazadjiev2014} employ the Einstein frame and only later the results are transformed to the physical Jordan frame. This was very surprising to us because there is a simple mathematical relation between the two frames that gives one to one correspondence in practically all of the physically relevant cases. The Einstein frame is used in a big part of the studies of compact objects in scalar-tensor and $f(R)$ theories since it simplifies the equations considerably. Moreover, the equation of state we use is in the Jordan frame, that rules out further ambiguity.

Our results in \cite{Yazadjiev2014,Staykov2014,Yazadjiev2015} were obtained independently with two different codes: a static code developed in \cite{Yazadjiev2014} and an extension to the {\tt RNS} code \cite{Stergioulas95} developed in \cite{Yazadjiev2015,Doneva2013} for the case of rapid rotation. The two codes use quite different forms of the metric and the field equations. The numerical methods in the two cases are also very different. All of this rule out possible errors. In order to further check our results and disprove the suggestion in \cite{Capozziello2015} that the difference in the results is due to the different frames, we have derived the reduced field equations in the Jordan frame and solved them numerically. As explained below, our results show perfect agreement with the previous studies \cite{Yazadjiev2014}. This proves  that the results in \cite{Yazadjiev2014} are correct and  shows that the numerical results in \cite{Capozziello2015} seem to be incorrect. Here we will also comment on the most probable source of error in \cite{Capozziello2015}.


\section{Basic equations}

The  $f(R)$ theories of gravity are defined by the action

\begin{eqnarray}\label{A}
S= \frac{1}{16\pi } \int d^4x \sqrt{-g} f(R) + S_{\rm
matter}(g_{\mu\nu}, \chi),
\end{eqnarray}
where $R$ is the scalar curvature with respect to the spacetime
metric $g_{\mu\nu}$ and $S_{\rm matter}$ is the action of the matter
fields denoted by $\chi$. Viable $f(R)$ theories have to be free
of tachyonic instabilities and the appearance of ghosts which
requires 

\begin{eqnarray}
\frac{d^2f}{dR^2}\ge 0,  \;\;\; \frac{df}{dR}>0,
\end{eqnarray}
respectively.

The field equations corresponding to the action (\ref{A}) are the following:

\begin{eqnarray}
f_{R} R_{\mu\nu} - \frac{1}{2} f g_{\mu\nu} - (\nabla_{\mu}\nabla_{\nu} - g_{\mu\nu}\square)f_{R}= 8\pi T_{\mu\nu},
\end{eqnarray}
where $T_{\mu\nu}$ is the matter energy-momentum tensor, $\nabla_{\mu}$ is the covariant derivative and $\square=g^{\mu\nu}\nabla_{\mu}\nabla_{\nu}$. These equations can also be
written in the following form:

\begin{eqnarray}\label{FE1}
f_{R}\left(R_{\mu\nu} - \frac{1}{2}R g_{\mu\nu}\right) - f_{RR}\nabla_{\mu}\nabla_{\nu}R - f_{RRR}\nabla_{\mu}R\nabla_{\nu}R + \left[\frac{1}{2}(Rf_{R} - f) + f_{RR} \square R +
f_{RRR} \nabla_{\sigma}R \nabla^{\sigma}R\right] g_{\mu\nu}= 8\pi T_{\mu\nu},
\end{eqnarray}
where $f_{R}$, $f_{RR}$ and $f_{RRR}$ are the first, the second and the third derivatives of the function $f(R)$ with respect to the Ricci scalar $R$.
The trace of the above equations gives

\begin{eqnarray}\label{FER}
\square R = \frac{1}{3f_{RR}}\left[8\pi T - 3 f_{RRR} \nabla_{\sigma}R\nabla^{\sigma}R + 2f - Rf_{R}\right].
\end{eqnarray}
Substituting back in (\ref{FE1}) we find

\begin{eqnarray}\label{FE2}
f_{R}\left(R_{\mu\nu} - \frac{1}{2}R g_{\mu\nu}\right)=8\pi T_{\mu\nu} + f_{RR}\nabla_{\mu}\nabla_{\nu}R +  f_{RRR}\nabla_{\mu}R\nabla_{\nu}R - \frac{1}{6}\left[f + Rf_{R} + 16\pi T\right] g_{\mu\nu}.
\end{eqnarray}
Equations (\ref{FER}) and (\ref{FE2}) are the basic field equations of the $f(R)$ theories of gravity.

The next step is to consider a static and spherically symmetric spacetime described by
the  metric

\begin{eqnarray}
ds^2 = - A(r)dt^2 + B(r)dr^2 + r^2 (d\theta^2 + \sin^2\theta d\phi^2).
\end{eqnarray}

Since the purpose of the present paper is to study the structure of neutron stars in
$f(R)$ gravity, we consider the matter source to be a perfect fluid with energy density $\rho$ and pressure $p$. We also require the perfect
fluid  to respect the staticity and the spherical symmetry. With these
conditions imposed, the dimensionally reduced field equations are

\begin{eqnarray}
&&A^{\prime} = \frac{A}{r(2f_{R} + r R^{\prime} f_{RR})} \left[Br^2(f-Rf_{R} + 16\pi p) + 2f_{R}(B-1) - 4rR^{\prime} f_{RR}\right],\\
&&B^{\prime}= \frac{B}{r(2f_{R} + r R^{\prime} f_{RR})} \left\{2f_{R}(1-B) + \frac{1}{3}(32\pi \rho + 48\pi p + Rf_{R} + f) Br^2
+\right. \nonumber \\ && \left.
\frac{rR^{\prime}f_{RR}}{f_{R}}\left[\frac{Br^2}{3}(2Rf_{R} - f + 16\pi\rho) + 2(1-B)f_{R} + 2rR^{\prime}f_{RR}\right] \right\} ,\\
&&R^{\prime\prime}= \frac{1}{3f_{RR}} \left[B(-8\pi \rho + 24\pi p + 2f - Rf_{R}) - 3 f_{RRR}{R^{\prime}}^2\right] + \left(\frac{B^{\prime}}{2B} -\frac{A^{\prime}}{2A} - \frac{2}{r} \right) R^{\prime}.
\end{eqnarray}

The contracted Bianchi identity $\nabla_{\mu}T^{\mu\nu}=0$ gives the equation for the hydrostatic equilibrium of the fluid,

\begin{eqnarray}\label{FEQE}
p^{\prime}= - (\rho + p)\frac{A^{\prime}}{2A}.
\end{eqnarray}

It has to be noted that the dimensionally reduced equations \eqref{eq:Field1}--\eqref{eq:Field3} and \eqref{FEQE} coincide with those derived in \cite{Jaime2011a}.

In what follows we will focus on a particular class of $f(R)$ theories of gravity, namely the $R^2$ gravity, where $f(R) = R + a R^2$ with $a\ge 0$ being a parameter. This case was considered both by \cite{Yazadjiev2014} and \cite{Capozziello2015} which facilitates the comparison. For $R^2$ gravity the field equations take the following form:

\begin{eqnarray}
&&A^{\prime}= \frac{A}{2r(1 + 2aR + arR^{\prime})}\left[Br^2(16\pi p -aR^2) + 2(1 +  2aR)(B-1) - 8arR^{\prime} \right],\label{eq:Field1}\\
&&B^{\prime}= \frac{B}{2r(1 + 2aR + arR^{\prime})}  \left\{ 2(1+2aR)(1-B)   + \frac{Br^2}{3}\left[2R + 3aR^2 + 16\pi(2\rho + 3p)\right]
\right. \nonumber \\ &&\left.+ \frac{2arR^{\prime}}{1+ 2aR}\left[\frac{Br^2}{3}(R + 3aR^2 + 16\pi\rho) + 2(1-B)(1+ 2aR) + 4arR^{\prime}\right] \right\},\label{eq:Field2}\\
&&R^{\prime\prime}= \frac{B}{6a}\left[R + 8\pi(3p-\rho)\right] + \left(\frac{B^{\prime}}{2B} - \frac{A^{\prime}}{2A} - \frac{2}{r}\right)R^{\prime}. \label{eq:Field3}
\end{eqnarray}

The boundary conditions come from the requirements of asymptotic flatness at infinity and regularity at the center of the star. Namely we impose
\begin{equation}
\left.\frac{dR}{dr}\right|_{r=0} = 0, \hskip 0.7cm B|_{r=0} = 1, \hskip 0.7cm R|_{r\rightarrow \infty} = 0, \hskip 0.7cm A|_{r\rightarrow \infty} = 1. \label{eq:BC}
\end{equation}

The radius of the star $R_s$ is determined as the point where the pressure vanishes, i.e. $p(R_s)=0$. The mass $M$ is calculated from the asymptotic behavior of the metric
\begin{equation}
M=\left.\frac{r}{2} \left(1 - \frac{1}{B}\right)\right|_{r\rightarrow\infty}. \label{eq:mass}
\end{equation}

In the presented numerical results we shall use the dimensionless parameter $a\to a/r^2_{0}$, where $r_{0}$ is one half of the solar gravitational radius   $r_{0}=1.47664 \,{\rm km}$ (i.e. the solar mass in geometrical units).

\section{Numerical results}
\begin{figure}[]
	\centering
	\includegraphics[width=0.45\textwidth]{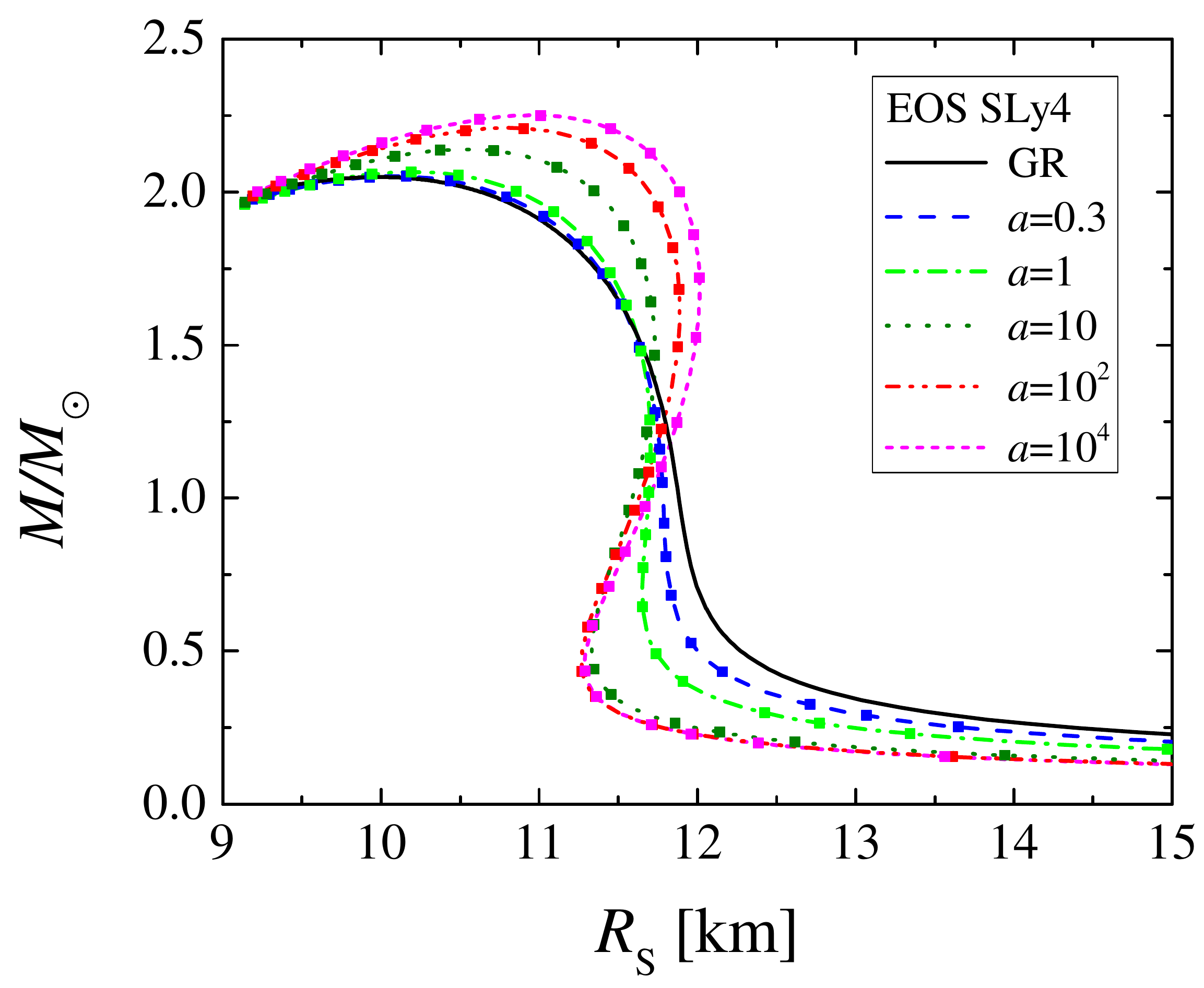}
	\includegraphics[width=0.45\textwidth]{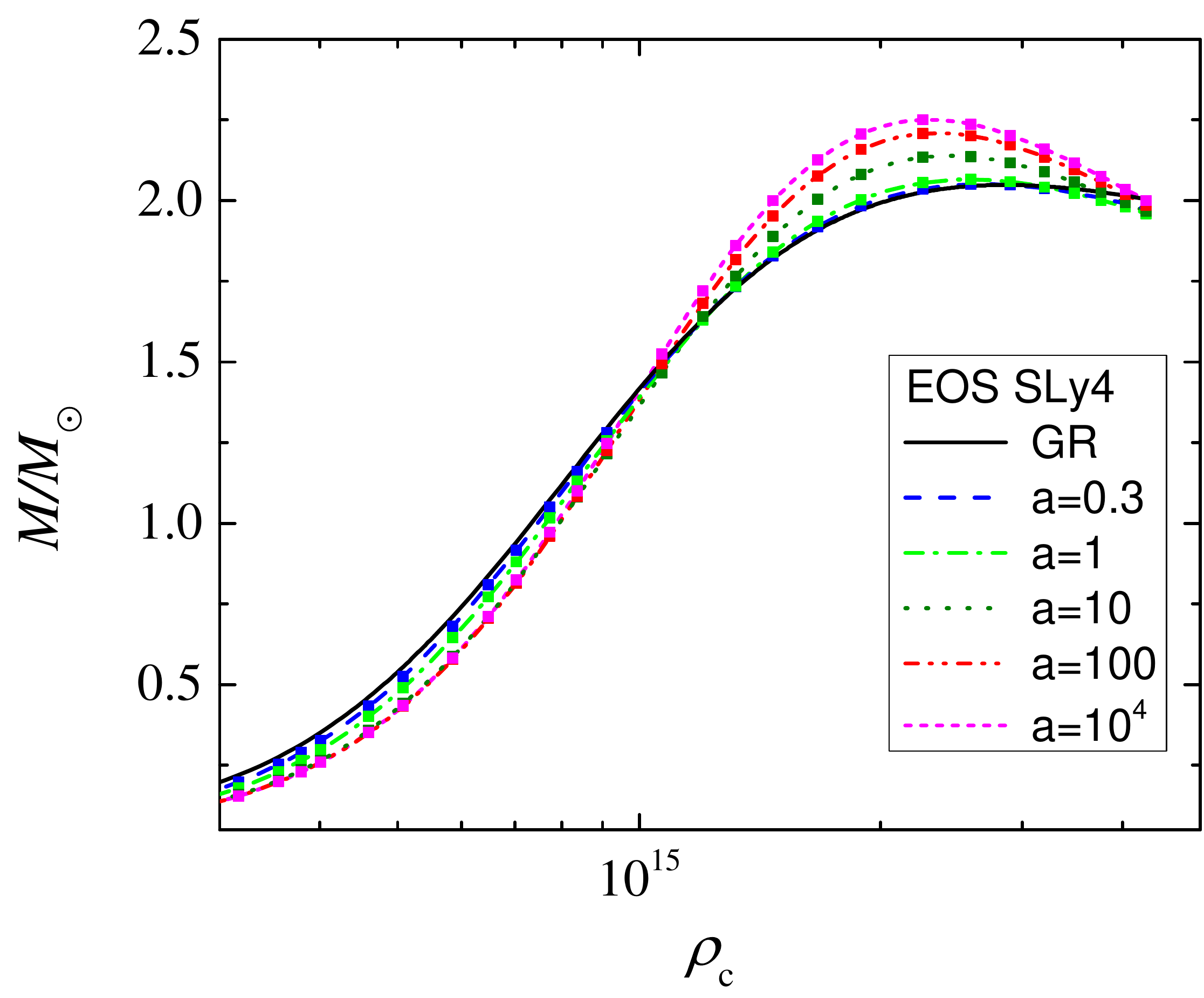}
	\caption{The mass of radius relation (left panel) and the mass as a function of the central energy density (right panel) for EOS SLy4. Different styles and colors correspond to different values of the parameter  $a$. The lines represent the results in the Einstein frame reported in \cite{Yazadjiev2014}, while the bullets are the data calculated in the Jordan frame.}
	\label{Fig:MR_AllEOS}
\end{figure}

We solve the reduced field equations \eqref{eq:Field1}--\eqref{eq:Field3} and \eqref{FEQE} with the corresponding boundary conditions \eqref{eq:BC} using a shooting method. The numerical results are presented in Fig. \ref{Fig:MR_AllEOS} for the case of equation of state (EOS) SLy4. We choose this particular EOS, because it was used in both \cite{Yazadjiev2014} and \cite{Capozziello2015}. The lines represent the results in the Einstein frame reported in \cite{Yazadjiev2015} while the bullets are the data calculated in the Jordan frame obtained when solving the field equations \eqref{eq:Field1}--\eqref{eq:Field3}. As one can see the two sets of data are in perfect agreement. Moreover, a detailed check of the values of the mass and the radius for models with the same central energy density and value of $a$ in the Einstein and the Jordan frame shows, that the difference between the results in the two frames is within the numerical error. This clearly shows once more that the two sets of equations (the one in the Jordan frame derived in the current paper and the Einstein frame equations in \cite{Yazadjiev2014}) are equivalent and lead to the same results.

The results in \cite{Capozziello2015} differ not only quantitatively but also qualitatively from the results in the present paper. For example with the increase of the parameter $a$ (or $|\alpha|$ in the notations of \cite{Capozziello2015}) the maximum mass starts to decrease, contrary to our results.\footnote{As a matter of fact a very slight decrease of the maximum mass (below 1 \%) was also observed in \cite{Yazadjiev2014}, typically for $a<1$. This is an almost marginal effect though and it clearly differs from the substantial decrease of the maximum mass, observed in \cite{Capozziello2015}.}

From  Fig. 1 presented in \cite{Capozziello2015} one can see that the scalar curvature inside and outside the star is positive. However, standard arguments similar to those in the maximum principle for elliptic differential equations show that the scalar curvature must be negative for the adopted metric signature in \cite{Capozziello2015}, for  negative parameter $\alpha$  and for the numerical value of the central density reported in the caption of the figure. This shows that the numerical results in \cite{Capozziello2015}, at least those for  $R^2$ gravity,  are incorrect.

The differences between our results and those obtained in \cite{Capozziello2015} could probably be explained  by the following mistakes we have noticed in \cite{Capozziello2015}, among many others:
\begin{enumerate}
\item The dimensionally reduced equation for the scalar curvature, namely eq. (14) in \cite{Capozziello2015}, is wrong as it is written. This equation does not give the right general relativistic limit for $\alpha\to 0$  in their notations. More precisely,  eq. (14) in \cite{Capozziello2015} does not recover their eq. (10) in the mentioned limit. The limit gives an expression with the wrong sign.
\item According to the authors of \cite{Capozziello2015} the dimensionless scalar curvature is $R/r^2_{0}$ (or in their notations $R/r^2_{g}$). This is a very strange claim. The dimensionless scalar curvature is $R r^2_{0}$ (or $R r^2_{g}$ in the notations of \cite{Capozziello2015}). This mistake could lead to a serious mess in the dimensionless equations that are solved in \cite{Capozziello2015}.
\end{enumerate}

We have also noticed differences between our dimensionally reduced equations and the dimensionally reduced field equations presented in \cite{Capozziello2015}.

\section{Conclusion}

In this comment we have addressed the observed large difference between the neutron star models in $f(R)$ theories presented in the recent paper \cite{Capozziello2015} and in our previous paper \cite{Yazadjiev2014}. These differences are not only quantitative but also qualitative. Our results in \cite{Yazadjiev2014} were extensively tested by two numerical codes using quite different forms of the field equations and the metric, which gives us the confidence to exclude a possible error. It was suggested in \cite{Capozziello2015} that the difference comes from the fact that they were using the physical Jordan frame while we were using the Einstein frame. We disagree with this statement because there is a simple mathematical relation between the two frames and furthermore we have used the EOS in the physical Jordan frame. In order to prove this, we have derived the field equations in the Jordan frame and solved them numerically. The results are in perfect agreement with \cite{Yazadjiev2014}. Therefore, the results in \cite{Capozziello2015} are most probably incorrect and we have pointed out several possible sources of error.

\acknowledgements{DD would like to thank the European Social Fund and the Ministry of Science, Research and the Arts Baden-W\"urttemberg for the support. The support by the Bulgarian NSF Grant DFNI T02/6 and "New-CompStar" COST Action MP1304 is gratefully acknowledged. }

\appendix

\bibliography{references}

\end{document}